\documentstyle[pra,aps,draft]{revtex}

\begin{document}

\title{Encoded Universality in Physical Implementations of a Quantum Computer}
\author{
D. Bacon,$^{1,2}$ 
J. Kempe,$^{1,3}$
D. P. DiVincenzo,$^{4}$
D. A. Lidar,$^{1}$\thanks{
Current address: Dept. of Chemistry, Univ.
of Toronto, Ontario, CANADA}
~and K. B. Whaley$^{1}$\thanks{
Corresponding author}}
\address{Departments of Chemistry$^1$, Physics$^2$ and
Mathematics$^3$, University of California, Berkeley 94270\\
$^4$IBM Research Division, T.J. Watson Research Center, Yorktown
Heights, New York 10598}

\maketitle
\abstract{
We revisit the question of universality in quantum computing and
propose a new paradigm. Instead of forcing a physical system to enact
a predetermined set of universal gates (e.g., single-qubit operations
and CNOT), we focus on the intrinsic ability of a system to act as a
universal quantum computer using only its naturally available interactions.  
A key element of this approach is the
realization that the fungible nature of quantum information allows for
universal manipulations using quantum information encoded in a
subspace of the full system Hilbert space, as an alternative to using physical
qubits directly. Starting with the interactions intrinsic to
the physical system, we show how to determine the
possible universality resulting from these interactions over an encoded 
subspace.
We outline a general Lie-algebraic framework that can be used to find the
encoding for universality, and give examples relevant to 
solid-state quantum computing.
}

\section{Introduction}

\label{intro}

Our generous universe comes equipped with the ability to compute. In
physics, determination of the \emph{allowable} manipulations of a physical
system is of central importance. Computer science, on the other hand, has
arisen in order to \emph{quantify} what resources are needed in order to
perform a certain algorithmic function. For computer science to be
applicable to the real world, this quantification should be limited by what
physics has determined to be allowable manipulations. Thus we arrive at the
realization that because information is physical, computer science should be
built on primitives which respect the laws of physics.

An important property of classical information which carries over to the
quantum world is the fungible nature of information. A resource is fungible
if interchanging it with another resource does not destroy the value of the
resource. Whether we represent a classical bit by the presence or absence of
a chad on a punch-card\cite{BushvsGore} or in the average orientation of a
million electron spins, the intrinsic value of the information (the value of
the bit) is untouched: it does not depend upon the medium in which it is
represented. The fungible nature of information has been key to the
exponential growth of the computer revolution: it does not matter that the
information is being confined to ever smaller components on silicon chips.
So, too, for quantum information: the plethora of experimentally proposed
systems from which a quantum computer could be built is made possible by the
fungibility of quantum information.

Another example of the fungible nature of quantum information is the idea
that one can \emph{encode} quantum information. The theory of quantum error
correcting codes, for instance, makes use of this to allow computation in
the presence of noise.\cite{Steane:99} In this paper we argue that we can
harness the fungible nature of quantum information in another important
aspect, namely in universality constructions. In particular, we propose that
instead of taking a physical system and adding external Hamiltonians and
interactions to it so that a certain universal set of gates may be
performed, one examines instead the intrinsic system interactions for their
potential universality. The key to this approach is the recognition that the
ability to encode quantum information allows for certain interactions to be
universal over an encoded Hilbert subspace, even though they are not
universal over the entire Hilbert space. Our task is then to search for the
encodings providing such universality, given the intrinsic system
interactions. Such encoded universality is common in quantum error
correction but appears to have not been recognized in the discussions
concerning the physical construction of a quantum computer.

\section{Universality}

\label{hist}

The action of gates $\mathbf{G}$ on a circuit corresponds to unitary
evolution of the system, with the gates given by the evolution operators
derived from the system Hamiltonian $\mathbf{H}$: $\mathbf{G}(t)=\mathtt{T}%
\left\{ \exp [-i\int^{t}\mathbf{H}(\tau )d\tau ]\right\} $. Here $\mathtt{T}$
denotes the time ordering operation. 
A
finite set $\mathcal{G}$ of gates is defined to be \emph{universal} if:
i) each $\mathbf{U}_{i}\in \mathcal{G}$ acts on a constant number of
qubits only, and
ii) any quantum circuit (performing any desired unitary operation) can
be simulated to arbitrary accuracy efficiently by a circuit with gates in $%
\mathcal{G}$.

Somewhat surprisingly, most discussions
of universality in quantum computing are cast in the language of gates,
rather than Hamiltonians. As physicists we find, however, that in order to
understand encoded universality, it is often more useful to think about a
gate in terms of the generating Hamiltonian instead of the resulting gate. 
Given the notion of a universal set of gates, various physical
implementations of such a quantum computer have been proposed. Naturally, in
order to qualify as a valid quantum computer, each physical implementation
must be shown to possess a universal set of gates arising from controllable
interactions of the given physical system. To date, with some 
exceptions,\cite{Bacon:99b,Kempe:00a,DiVincenzo:00a} all physical proposals for a
quantum computer have implemented a set of universal gates via a route in
which: (i) the basic qubit is identified in the physical system, (ii)
single-qubit unitary gates are shown to be possible on these qubits, and
(iii) some two-qubit gate acting between qubits is shown to be possible. The
latter is usually a CNOT or a controlled phase. For a few systems this
standard model does provide a natural path towards building a quantum
computer, but for other implementations severe device engineering must be
performed in order to force the system to achieve this universal set of
gates.

For instance, in many solid state implementations a fast intrinsic
interaction is the exchange interaction (Heisenberg coupling) 
\begin{equation}\label{eq:Eij}
\mathbf{E}_{ij}=\vec{\sigma}^{i}\cdot \vec{\sigma}^{j},  
\end{equation}
(here $\vec{\sigma}^{i}$ is the vector of Pauli spin matrices $({\sigma _{x}}%
,{\sigma }_{y},{\sigma }_{z})$ on the $i$-th qubit). We call $\mathbf{E}_{ij}
$ the exchange interaction because addition of a global phase and rescaling
gives a Hamiltonian which exchanges the qubits $(|x\rangle |y\rangle
\rightarrow |y\rangle |x\rangle )$.\cite{Lidar:00} The gates which can be
obtained by tuning the exchange interaction strength $J$ are the unitary
gates $\mathbf{U}_{ij}(t)=\exp (itJ\mathbf{E}_{ij})$. Given $n$ qubits
coupled via exchange interactions it can be shown\cite{Kempe:00a} that these
gates alone cannot be used to implement an arbitrary quantum circuit on $n$
qubits. Thus designers of solid state quantum computers have had to
supplement the exchange interaction by single-qubit rotations in order to
make their set of interactions universal over all $n$ qubits. However,
adding this single-qubit capability generally leads to considerable device
complexity and to disadvantages such as slow single-qubit gate times and
possible additional sources of decoherence resulting from addition of
auxiliary, non-intrinsic interactions to the system.

It was recently realized\cite{Bacon:99b,Kempe:00a,DiVincenzo:00a} that the
exchange interaction by itself can be used to simulate a quantum circuit if
a particular encoded space of three or more qubits is used to represent the
quantum information. Thus while the exchange interaction on $n$ qubits
cannot be used to implement an arbitrary quantum circuit on $n$ qubits, it
can be used to simulate a quantum circuit on $cn$ qubits, where $c$ is a
constant representing the encoding efficiency of such a code. The fact that
the exchange interaction \textit{alone} can be used to simulate a quantum
circuit by using encoded quantum information allows one to avoid
implementing the generally difficult single-qubit gates.

This example illustrates a new approach to universality very different from
the conventional paradigm of using single qubit-gates supplemented by a
two-qubit gate. The latter has consituted the primary guideline in recent
years for deciding whether a physical system might be used as a quantum
computer, without regard to the fact that this is just one of many possible
paths to universality. Nevertheless, other approaches have also been
proposed. Thus, it was shown\cite{Deutsch:95,Lloyd:95a} that \emph{almost any%
} two-qubit unitary gate along with the exchange gate acting between $n$
qubits is universal over all $n$ qubits. 
We propose here a radically different route to achieve universality, namely,
that starting from the natural interactions given by the physics of the
proposed qubit system (e.g., $\mathbf{E}_{ij}$), the underlying potential of
that interaction for encoded universality be investigated. For a given
physical system, the natural interactions of choice should be determined by
various factors including their speed, the ease with which they can be
implemented, and their robustness towards decoherence processes. The path we
take, then, is to proceed from the intrinsic interactions to encoded
universality, rather than to artificially modify or build the interactions
to fit a standard model.

\section{General Formalism}

\label{general}

The general formalism will be illustrated here with the example of the
exchange interaction. We omit most of the proofs and details, which can be
found in Ref. 4.

Assume that we are given a set of interactions on some $d$-dimensional
Hilbert space -- Hermitian Hamiltonians $h_{i}(t)\mathbf{H}_{i}$ where $%
h_{i}(t)$ are controllable coupling constants -- that we want to use for our
universal quantum computer. Which unitary gates can be approximated by
turning these interactions on and off (perhaps in parallel)? The answer to
this\cite{Deutsch:95,Lloyd:95a,DiVincenzo:95a} is that every unitary gate in
the Lie group corresponding to the Lie algebra \emph{generated} by the
Hamiltonians $\mathbf{H}_{i}$ can be approximated to arbitrary accuracy by a
sequence of gates obtained through the temporal evolution of $h_{i}(t)%
\mathbf{H}_{i}$. This is true because of the following properties: 
\begin{eqnarray}
\label{Trotter}
&&e^{i(\alpha \mathbf{A}+\beta \mathbf{B})}=\lim_{n\rightarrow \infty
}(e^{i\alpha \mathbf{A}/n}e^{i\beta \mathbf{B}/n})^{n}  \label{Comm} \\
&&e^{[\mathbf{A},\mathbf{B}]}=\lim_{n\rightarrow \infty }(e^{-i\mathbf{A}
\sqrt{n}}e^{i\mathbf{B}/\sqrt{n}}e^{i\mathbf{A}/\sqrt{n}}e^{-i\mathbf{B}
\sqrt{n}})^{n}.
\end{eqnarray}
To approximate to a given accuracy, say $\exp (i(\alpha \mathbf{A}+\beta 
\mathbf{B}))$, by a sequence of gates with generators $\mathbf{A}$ and $%
\mathbf{B}$ one just truncates Eq.~(\ref{Trotter}) for large enough $n$.
Every element in the Lie algebra, that is the algebra formed via linear
combination ($\alpha \mathbf{A}+\beta \mathbf{B}$) and via the Lie
commutator ($i[\mathbf{A},\mathbf{B}]$), can be approximated to given
accuracy using these results. Thus all gates in the Lie group corresponding
to this Lie algebra generated by the $\mathbf{H}_{i}$ can be approximated to
arbitrary accuracy by using the controllable coupling constants $h_{i}(t)$.
The inverse of this statement is also true.

Let $\mathcal{A}$ be the Lie algebra generated by the $\mathbf{H}_{i}$. The
Lie group generated by the Lie algebra $\mathcal{A}$ is a subgroup of the $d$%
-dimensional unitary group $U(d)$ and thus is a compact Lie group. An
important property of compact Lie groups is that they, along with the Lie
algebra which generates the Lie group, are completely reducible. A
representation of a Lie algebra is irreducible if the action of the
representation on its vector space does not possess an invariant subspace.
An algebra which is completely reducible can be written as a direct sum of
irreducible representations (irreps): 
\begin{equation}
{\mathcal{A}}\cong \bigoplus_{J}\bigoplus_{\lambda =1}^{n_{J}}{\mathcal{L}}_{J},
\end{equation}
where ${\mathcal{L}}_{J}$ is the $J$th irrep which appears with degeneracy $%
n_{J}$ and is $d_{J}$-dimensional. For the operators $\mathbf{H}_{i}$ this
implies that there is a basis for the Hilbert space on which $\mathbf{H}%
_{i}$ acts as
\begin{equation}
\mathbf{H}_{i}=\bigoplus_{J}\mathbf{I}_{n_{J}}\otimes (\mathbf{L}_{J})_{i}
\label{eq:decomp}
\end{equation}
where $(\mathbf{L}_{J})_{i}$ are elements of the $J$th irrep ${\mathcal{L}}%
_{J} $ and $\mathbf{I}_{d}$ is the $d$-dimensional identity operator. 
Clearly, understanding the irreps of the Lie algebra generated by the
Hamiltonians $\mathbf{H}_{i}$ will allow us to understand the suitability of
this Lie algebra for simulating a quantum circuit. In particular, Eq.~(\ref
{eq:decomp}) implies that with a \emph{suitable encoding} the $\mathbf{H}_{i}$ can produce the action of a specific irrep ${\mathcal{L}}_{J}$. 
This action may or may not be useful for quantum
computation, an issue we address below. 

It is often helpful in understanding the universality of an interaction to
find the \emph{commutant} ${\mathcal{A}}^{\prime }=\left\{ \mathbf{X}:[\mathbf{X},\mathbf{A}]=0,\forall \mathbf{A}%
\in {\mathcal{A}}\right\} $ of the Lie algebra $%
{\mathcal{A}}$.
It should be clear from Eq.~(\ref{eq:decomp}) that every element $\mathbf{K}\in \mathcal{A%
}^{\prime }$ is reducible to: 
\begin{equation}
\mathbf{K}=\bigoplus_{J}{\mathcal{M}}_{n_{J}}\otimes \mathbf{I}_{d_{J}}
\label{eq:decompA'}
\end{equation}
where ${\mathcal{M}}_{n_{J}}$ is an $n_{J}$-dimensional (complex) square
matrix, and $d_{J}$ is the dimension of the $J$th irrep. This structure, dual to $\mathcal{A}$,
offers a useful way for seeing how the encoding arises: $\mathcal{A}^{\prime }$
splits into a sum of degenerate irreps.
The degeneracy of a particular
representation gives the dimension of the space over which the encoding is
made. 

We should point out a non-trivial aspect of the question of ``universality
given a set of interactions'', which makes our task more difficult. This is
the tensor product nature of quantum computation. We recall that a set of
interactions is universal if it can be used to efficiently simulate a
quantum circuit. An important property of the quantum circuit model is that
at its most basic level it possess a tensor product of single qubit systems.
Hidden within this seemingly trivial fact is the notion of locality: we
believe that physics is local and hence we require our model of computation
based on physics to preserve this locality. For the purposes of building a
universal quantum computer this implies that we must, at some point,
introduce a suitable tensor product structure in order to \emph{efficiently}
simulate a quantum circuit. The above discussion of representation theory
for Lie algebras has been solely in terms of some abstract $d$-dimensional
Hilbert space. We now admit two tensor product structures: first, on our
physical system, and second, on the encoded qubits from which encoded
universality will be constructed. The first tensor product structure is
simply that implied by our physical system and is forced on us by the
locality of physics. For example, if we are using the spins of single
electrons on a quantum dot, the tensor product is just the natural one of
these spin-qubits. The nature of this tensor product structure is also
manifest in the set of interactions which will be present in the real world,
and is represented in the Hamiltonian as operators acting on separate
degrees of freedom.

The second tensor product is required by the fact that we are going to
simulate a quantum circuit. Given a set of interactions, we know that these
can be used to produce the action of a given set of irreps ${\mathcal{L}}_{J}$%
. However, without a mapping from the subspace on which the irrep acts to
the tensor product structure of a quantum circuit, this is not enough to
quantify the suitability of the interactions for universality. The exact
cutoff where this tensor product structure comes into play is somewhat
arbitrary. We will take the position here that we are searching for \emph{%
small encodings}. That is, we seek to minimize the number of physical qubits
used to encode a logical qubit. This logical qubit will be formed by a block
of nearest neighbor qubits with the defining property that single
encoded-qubit operations are possible within each such block. Given a set of
such encoded qubits we then form a tensor product between pairs of blocks.
We refer to this process of forming a tensor product code as ``conjoining''.
These considerations are illustrated below.

\section{Example of Isotropic Exchange}

\label{examples}

As an example of our general discussion consider the Heisenberg model for
spin-spin interactions. The most general form of this is given by a sum of
fully anisotropic pair-wise spin-spin couplings 
\begin{equation}
\mathbf{H}_{ij}=J_{ij}^{X}{\sigma }_{x}^{i}{\sigma }_{x}^{j}+J_{ij}^{Y}{%
\sigma }_{y}^{i}{\sigma }_{y}^{j}+J_{ij}^{Z}{\sigma }_{z}^{i}{\sigma }%
_{z}^{j}.  \label{eq:hei_anis}
\end{equation}
We consider in detail here the isotropic limit in which the exchange
couplings between a given $ij$ pair are all equal, \textit{i.e.}, $%
J_{ij}^{X}=J_{ij}^{Y}=J_{ij}^{Z}\equiv J_{ij}$, so that we are dealing then
with the Heisenberg Hamiltonian 
\begin{equation}
\mathbf{H}_{\mathrm{Hei}}=\sum_{i\neq j}J_{ij}\mathbf{E}_{ij}
\label{eq:hei_iso}
\end{equation}
where we assume that the coupling parameters $J_{ij}$ can be turned on/off
at will. To show how this intrinsic coupling can lead to encoded
universality we analyze its action on three qubits.

The Lie algebra ${\mathcal{L}}_{E}$ generated by $\mathbf{E}_{ij}$ on three
qubits allows us to implement the Hamiltonians in the set $\{\mathbf{E}_{12},%
\mathbf{E}_{23},\mathbf{E}_{13},\mathbf{T}\equiv i([\mathbf{E}_{12},\mathbf{E%
}_{23}]\}$. A better basis for the Lie algebra is given by the set of
operators $\mathbf{H}_{0}=\mathbf{E}_{12}+\mathbf{E}_{23}+\mathbf{E}_{13}$, $%
\mathbf{H}_{1}={\frac{1}{4\sqrt{3}}}\left( \mathbf{E}_{13}-\mathbf{E}%
_{23}\right) $, $\mathbf{H}_{3}={\frac{1}{12}}\left( -2\mathbf{E}_{12}+%
\mathbf{E}_{23}+\mathbf{E}_{13}\right) $, $\mathbf{H}_{2}=i[\mathbf{H}_{3},%
\mathbf{H}_{1}]$. We then find that 
\begin{equation}
\lbrack \mathbf{H}_{0},\mathbf{H}_{\alpha }]=0
\end{equation}
for all $\alpha $ and 
\begin{equation}
\lbrack \mathbf{H}_{\alpha },\mathbf{H}_{\beta }]=i\epsilon _{\alpha \beta
\gamma }\mathbf{H}_{\gamma }
\end{equation}
with $\alpha ,\beta ,\gamma \in \{1,2,3\}$. $\mathbf{H}_{0}$ is an abelian
invariant subalgebra of this Lie algebra and thus factors out as a global
phase. The set $\{\mathbf{H}_{1},\mathbf{H}_{2},\mathbf{H}_{3}\}$, on the
other hand, act as the generators of $su(2)$. Thus the exchange interaction
between three qubits can be used to implement a single encoded-qubit $su(2)$%
. More precisely, we find that the $\mathbf{H}_{\alpha }$ for $\alpha \in
\{1,2,3\}$ generates the algebra 
\begin{equation}
{\mathcal{L}}_{E}^{(3)}=\left( \bigoplus_{i=1}^{4}{\mathcal{S}}_{1}\right)
\oplus \left( \bigoplus_{i=1}^{2}{\mathcal{S}}_{2}\right) 
\end{equation}
where ${\mathcal{S}}_{d}$ is the $d$-dimensional irrep of $su(2)$. Note the
degeneracy of the corresponding irreps. Corresponding to this decomposition
the $\mathbf{H}_{\alpha }$ act as 
\begin{equation}
\mathbf{H}_{\alpha }=\mathbf{0}_{4}\oplus \left( {\frac{1}{2}}{\sigma }%
_{\alpha }\otimes \mathbf{I}_{2}\right) 
\end{equation}
for $\alpha \in \{1,2,3\}$ where $\mathbf{0}_{4}$ is 4-dimensional zero
operator (the $1$ dimensional irreps all act as $0$) and $\mathbf{I}_{2}$ is
the two-dimensional identity operator. The action of the exchange is thus
identical to that of an $su(2)$ operator on a single qubit when working over
the encoded space defined by the above decomposition. If we encode our
logical qubits as 
\begin{eqnarray}
|0_{L}\rangle  &=&{\frac{1}{\sqrt{2}}}(|010\rangle -|100\rangle )  \nonumber
\\
|1_{L}\rangle  &=&\sqrt{\frac{2}{3}}|001\rangle -\sqrt{\frac{1}{6}}%
|010\rangle -\sqrt{\frac{1}{6}}|100\rangle 
\end{eqnarray}
then we find that the action of $\mathbf{H}_{\alpha }$ is ${\frac{1}{2}}{%
\sigma }_{\alpha }$. Due to the degeneracy of the irrep, other encodings are
also possible. Note that these states are nothing but $J=1/2$ total angular
momentum states of $3$ spin-$1/2$ particles with a given projection along a
certain axis, and can thus be found using elementary addition of angular
momentum.\cite{Kempe:00a}

Having shown that the action of the exchange interaction on three qubits can
produce the effect of a single $2$-dimensional representation of $su(2)$, it
is natural to induce a tensor product structure between blocks of three
qubits in order to simulate a quantum circuit. We thus \emph{choose} an
encoding scheme in which a single qubit is identified with $3$ physical
qubis. This is not the only choice of tensor product structure. In fact, any
tensor product between sets of $k\geq 3$ qubits can be used to construct a
universal gate set. We define the encoding efficiency of our code as the
number of encoded bits of our simulated quantum computer ($e=\log _{2}(d)$
where $d$ is the dimension of the encoded information) divided by the total
number of qubits on which the encoding exists: $E=e/n=\log _{2}(d)/n$. Thus,
for a scheme using the exchange interaction between three qubits, where we
have a $d=2$ dimensional encoding on $k=3$ qubits, we find that $E=1/3$. A
peculiar property of the exchange interaction is that it has an asymptotic
encoding efficiency of unity. In particular, if we take $n$ qubits and use
the exchange interaction to compute on these qubits then the dimension of
the encoded space on which we can compute scales in such a way that the
encoding efficiency is at least $E(n)=1-\frac{3}{2}{\frac{\log _{2}n}{n}}$.
\cite{Kempe:00a} As $\lim_{n\rightarrow \infty }E(n)=1$ we loose almost
nothing in terms of encoding efficiency when we use the exchange
interaction. We note, however, that one must still introduce a tensor
product structure at some level in order to achieve universal quantum
computation.

Once one has introduced a tensor structure for the encoding, it is necessary
to show that the natural couplings of the system can produce a non-trivial
action between the tensor components of this encoded tensor product
structure. In the above case, where we have shown that we can obtain full
control over an encoded qubit, what we now need to show is that a
non-trivial action between the encoded qubits can be enacted. This is
nothing more than a map from the encoded universality to the fully universal
set of physical gates consisting of single qubit gates supplemented by a
non-trivial coupling between the qubits, i.e., to the standard model. This
point of the universality proof is typically the most daunting, but really
amounts to nothing more than understanding the Lie algebra generated by the
interaction over two tensored encoded qubits. For the exchange interaction
it can be shown that nearest neighbor exchange interactions can be used to
produce such an action.\cite{Kempe:00a} In particular, we find that the
effect of the exchange interaction on $6$ qubits becomes 
\begin{equation}
{\mathcal{L}}_{E}^{(6)}=\left( \bigoplus_{i=1}^{7}{\mathcal{S}}_{1}\right)
\oplus \left( \bigoplus_{i=1}^{5}{\mathcal{S}}_{5}\right) \oplus \left(
\bigoplus_{i=1}^{3}{\mathcal{S}}_{9}\right) \oplus {\mathcal{S}}_{5}
\end{equation}
When we conjoin the two three-qubit codes, we find that these codes lie
entirely within the last two irreps of this decomposition. This in turn
implies that non-trivial interactions between the encoded irreps are
possible.\cite{Kempe:00a} In fact the original single encoded qubit $su(2)$
is also contained within this decomposition. The important point, however,
is that on the tensor product between encoded qubits we can indeed couple
these encoded qubits in such a way as to achieve a map to the unencoded,
fully universal set of gates. Thus we have established that our encoding can
efficiently (to within a factor of three in spatial resources) simulate the
unencoded set of universal gates.

A similar analysis can be applied to the anisotropic Heisenberg model (XY
model). There it turns out that the smallest encoding one can achieve is a
logical qutrit (3-level system) into three physical qubits.\cite{tbp} For
both the isotropic and anisotropic Heisenberg models these universality
results are general, i.e., they can be shown to hold for arbitrary numbers
of qubits, $n$.\cite{Kempe:00a,tbp} However, the approach of conjoining small
blocks that encode qubits via a tensor product can be generally used to
construct the required mapping of universality for arbitrary interactions,
even when a general proof appears inaccessible.

\section{Outlook}

\label{outlook}

We have shown that the commonly assumed paradigm of universal quantum
computation requiring a set of both single-qubit and two-qubit gates can be
replaced by a more general notion of encoded universality. This approach
allows universality of quantum computation to be achieved via encodings
specific to the natural or convenient physical interactions in a system. In
the example given here we have discussed an instance where encoded
universality can be achieved using only a \textit{single} underlying
physical interaction, namely the exchange interaction. This result is of
particular significance for solid state implementations of quantum
computation. The approach of coupling blocks containing small numbers of
physical qubits (each forming one encoded qubit) via tensor products appears
to be applicable to any interaction.

The results for encoded universality that we have described here are
primarily existential, and do not address the very practical concerns of the
overhead in temporal and spatial resources which may result from encoding.
The practical question is then: what is the trade-off in resources of time
(gates) and space (qubits) resulting from the universality encoding for a
given set of physical interactions? Finding a universality encoding
generally immediately reveals the spatial overhead, \textit{i.e.}, the
number of physical qubits required to perform the encoding, and may also
provide information about the asymptotic overhead. Thus, for the exchange
interaction, the maximum spatial overhead is a factor of three, and becomes
arbitrarily small asymptotically.\cite{DiVincenzo:00a} The overhead in
temporal resources, \textit{i.e.}, how many physical gates are required to
implement a given set of encoded universal gates, is somewhat more
challenging to deal with. General issues of optimization of physical gates
for circuits of encoded qubits have been addressed.\cite{Burkard:99} It has
recently been shown explicitly that the encoded universal operations
deriving from the exchange interaction can be realised within $\sim $ 10
clock cycles.\cite{DiVincenzo:00a} However there does not exist to our
knowledge any general methodology to find the optimal gate sequence for a
specific two-qubit (or two-qutrit) gate.

The finding of encoded universality for the Heisenberg Hamiltonians suggest
that this approach will be very useful for consideration of solid state
quantum computation. 
Establishment
of results concerning the existence of encodings for general classes of
Hamiltonians is therefore now of interest. The
new paradigm of encoded universality clearly opens up significant new
opportunities for exploring the architectural realization of quantum
computation.

Acknowledgements:- 
This work was supported in full by the National Security Agency (NSA) and 
Advanced Research and Development Activity (ARDA) under Army Research 
Office (ARO) contracts DAAG55-98-1-0371 (KBW) and DAAG55-98-C-0041 (DPD).

\end{document}